\documentclass[english,jpclcd,article]{achemso}
\usepackage[T1]{fontenc}
\usepackage[latin9]{inputenc}
\usepackage{babel}
\usepackage{amsbsy}
\usepackage{amstext}
\usepackage{amssymb}
\usepackage{graphicx}
\usepackage{esint}
\usepackage[unicode=true,pdfusetitle,
 bookmarks=true,bookmarksnumbered=false,bookmarksopen=false,
 breaklinks=false,pdfborder={0 0 1},backref=false,colorlinks=false]
 {hyperref}
\usepackage{breakurl}

\makeatletter

\title{Predictive Semiclassical Model for Coherent and Incoherent Emission
in the Strong Field Regime: The Mollow Triplet Revisited}
\author{Hsing-Ta Chen}
\email{hsingc@sas.upenn.edu}
\affiliation{Department of Chemistry, University of Pennsylvania, Philadelphia,
Pennsylvania 19104, U.S.A.}
\author{Tao E. Li}
\affiliation{Department of Chemistry, University of Pennsylvania, Philadelphia,
Pennsylvania 19104, U.S.A.}
\author{Abraham Nitzan}
\affiliation{Department of Chemistry, University of Pennsylvania, Philadelphia,
Pennsylvania 19104, U.S.A.}
\author{Joseph E. Subotnik}
\affiliation{Department of Chemistry, University of Pennsylvania, Philadelphia,
Pennsylvania 19104, U.S.A.}
\newcommand{\lyxdot}{.}

\setkeys{acs}{articletitle = true}
\newcommand{\onlinecite}[1]{\hspace{-1 ex} \nocite{#1}\citenum{#1}} 
\newlength{\dhatheight}

\makeatother

\begin{document}
\begin{abstract}
We re-investigate the famous Mollow triplet and show that most of
the well-known quantum characteristics of the Mollow triplet\textemdash including
incoherent emission and a non-standard dependence of the sidebands
on detuning\textemdash can be recovered quantitatively using semiclassical
dynamics with a classical light field. In fact, by not relying on
the rotating wave approximation, a semiclassical model predicts some
quantum effects beyond the quantum optical Bloch equation, including
higher order scattering and asymmetric sideband features. This letter
highlights the fact that, with strong intensities, many putatively
quantum features of light\textendash matter interactions arise from
a simple balance of mean-field electrodynamics and elementary spontaneous
emission which requires minimal computational cost. Our results suggest
that the application of semiclassical electrodynamics to problems
with strong light\textendash matter coupling in the fields of nanophotonics
and superradiance are likely to yield a plethora of new information.
\end{abstract}

\newpage{}

Light\textendash matter interactions are interesting for two reasons:
1. Light scattered from matter carries detailed information about
the matter itself; 2. Generating quantum light sources and entangled
light\textendash matter states is crucial for novel applications in
quantum technology. For example, via light\textendash matter interactions,
a beam of classical light (a coherent laser) can be transformed into
non-classical light (photons) with quantum features, such as fluctuations
and cascade emission. In this letter, we will focus on one of the
simplest examples of matter interacting with a strong light field:
a two-level system (TLS) strongly driven by a monochromatic light.
For such a case, it is well known that the TLS will emit light with
a frequency spectrum composed of three peaks, known as a \emph{Mollow
triplet}.\cite{newstein_spontaneous_1968,mollow_power_1969}

The Mollow triplet spectrum is a universal signature of a quantum
system undergoing resonance fluorescence and cannot be explained by
classical optics: the prediction of the Mollow triplet spectrum is
one of the key achievement of \emph{quantum} optics. The salient
features of the Mollow triplet are as follows. First, the spectrum
is composed of one central peak at the driving frequency and two sidebands
shifted from the central peak by the Rabi frequency.\cite{kimble_theory_1976,del_valle_mollow_2010,moelbjerg_resonance_2012,mccutcheon_model_2013}
Second, at resonance, the fluorescence triplet is dominated by incoherent
emission.\cite{mollow_power_1969,cohen-tannoudji_dressed-atom_1977,ates_post-selected_2009,hughes_anisotropy-induced_2017}
Third, a radiative cascade leads to photon antibunching and blinking.\cite{aspect_time_1980,schrama_intensity_1992,nienhuis_spectral_1993,lopez_carreno_exciting_2015,kimble_photon_1977}
These three universal quantum optical features have been demonstrated
in (artificial) atomic systems,\cite{schuda_observation_1974,wu_investigation_1975,wrigge_efficient_2008,astafiev_resonance_2010,schulte_quadrature_2015}
quantum dots,\cite{xu_coherent_2007,muller_resonance_2007,vamivakas_spin-resolved_2009,flagg_resonantly_2009,ulrich_dephasing_2011,ulhaq_cascaded_2012,ulhaq_detuning-dependent_2013,lagoudakis_observation_2017}
and nitrogen vacancy defects.\cite{rohr_synchronizing_2014} 

To understand the features above, the usual approach in quantum optics
is to propagate the quantum optical Bloch equation (OBE), and then
infer the scattered field from the correlation function of the matter
using quantum electrodynamics (QED).\cite{cohen-tannoudji_photons_1997,cohen-tannoudji_atom-photon_1998,salam_molecular_2010}
More precisely, one makes an independent process assumption:\cite{cohen-tannoudji_atom-photon_1998}
the TLS responds to the incident field (a single, highly occupied
photon mode) and relaxes as the radiation field (a reservoir of unoccupied
photon modes) gains population by fluorescence. The propagation of
the quantum OBE usually relies on the rotating wave approximation
(RWA), which is valid near resonance and allows for analytical expressions.\cite{cohen-tannoudji_dressed-atom_1977}
Note that, however, the RWA solution accounts for the dynamics involving
only one resonant photon absorption/emission transition.

In this letter, we show that, quite surprisingly, most of the quantum
features of the Mollow triplet can be captured by semiclassical simulations
without relying on the RWA. In fact, a semiclassical model can predict
some quantum effects beyond what is possible with the OBE and the
RWA, such as higher order sidebands and an asymmetric spectrum. Furthermore,
because classical electrodynamics is so inexpensive, the methods discussed
below should be applicable to large systems in environments with arbitrary
dielectrics where a rigorous quantum treatment is not feasible nowadays.

\paragraph*{Driven Quantum System\label{par:Driven-Quantum-System}}

To simulate the features of the Mollow triplet, we consider a TLS
driven by a single-mode laser field in an effectively 1D space. The
electronic TLS Hamiltonian is $\widehat{H}_{S}=\hbar\omega_{0}\left|1\right\rangle \left\langle 1\right|$
where the ground state $\left|0\right\rangle $ and the excited state
$\left|1\right\rangle $ are separated by an energy difference $\hbar\omega_{0}$.
The transition dipole moment operator is given by $\widehat{\boldsymbol{{\cal P}}}\left(x\right)=\mathbf{d}\left(x\right)\left(\left|0\right\rangle \left\langle 1\right|+\left|1\right\rangle \left\langle 0\right|\right)$.
Here, for simplicity, we assume that the polarization distribution
takes the form of $\mathbf{d}\left(x\right)=\mu\sqrt{\frac{a}{\pi}}e^{-ax^{2}}\mathbf{e}_{z}$
and has a uniform charge distributions in the $yz$ plane. We also
assume that the width of the polarization distribution ($\sigma=1/\sqrt{2a}$)
is chosen to be relatively small in space so that the long-wavelength
approximation is valid. The quantum state of the TLS is described
by either the electronic density matrix $\widehat{\rho}$ or an ensemble
of electronic wavefunctions $\{\left|\psi^{\ell}\right\rangle $\}.

For this semiclassical model, both the incident laser photons and
the scattered field are treated as classical electromagnetic (EM)
fields. We assume that the incident laser photons are in a coherent
state, so that the average field observables closely resemble a continuous
wave (CW) EM field. Thus, for simplicity, the incident EM field takes
the form: $\mathbf{E}_{L}\left(x,t\right)=\frac{A_{L}}{\sqrt{\epsilon_{0}}}\cos\omega_{L}\left(\frac{x}{c}-t\right)\mathbf{e}_{z}$
and $\mathbf{B}_{L}\left(x,t\right)=-\sqrt{\mu_{0}}A_{L}\cos\omega_{L}\left(\frac{x}{c}-t\right)\mathbf{e}_{y}$
where $A_{L}$ is the amplitude and $\omega_{L}$ is the frequency
of the incident laser field. As $\sigma\rightarrow0$, the driving
term can be approximated by $-\int\mathrm{d}x\mathbf{E}_{L}\left(x,t\right)\cdot\mathbf{d}\left(x\right)\approx\hbar\Omega_{r}\cos(\omega_{L}t)$
where $\Omega_{r}=-\mu A_{L}/\hbar$ is the Rabi frequency. The CW
fields $\mathbf{E}_{L}$ and $\mathbf{B}_{L}$ satisfy source-less
Maxwell equations: $\frac{\partial}{\partial t}\mathbf{B}_{L}=-\boldsymbol{\nabla}\times\mathbf{E}_{L}$,
$\frac{\partial}{\partial t}\mathbf{E}_{L}=c^{2}\boldsymbol{\nabla}\times\mathbf{B}_{L}$.
In other words, the CW fields constitute standalone external fields
driving the TLS. Let us now describe how we model the dynamics of
the TLS state plus the scattered EM fields.

\paragraph*{Ehrenfest+R Dynamics for Driven TLS\label{sec:Ehrenfest+R-Dynamics}}

To model the Mollow triplet, we use ``Ehrenfest+R'' dynamics\textemdash a
simple amended version of Ehrenfest dynamics to account for spontaneous
emission.\cite{chen_ehrenfest+r_2019} In the context of a driven
TLS, the fundamental variables for Ehrenfest+R dynamics are the same
as those for Ehrenfest dynamics: $\left\{ c_{0}^{\ell},c_{1}^{\ell},\mathbf{E}_{S}^{\ell},\mathbf{B}_{S}^{\ell}\right\} $
where we use $\ell$ to index each trajectory. Recall that the coupled
equations of motion for Ehrenfest dynamics are of the standard Maxwell\textendash Schrodinger
form:
\begin{eqnarray}
\frac{\partial}{\partial t}c_{0}^{\ell}\left(t\right) & = & \frac{i}{\hbar}H_{01}\left(t\right)c_{1}^{\ell}\left(t\right),\label{eq:c0}\\
\frac{\partial}{\partial t}c_{1}^{\ell}\left(t\right) & = & -i\omega_{0}c_{1}^{\ell}\left(t\right)+\frac{i}{\hbar}H_{01}\left(t\right)c_{0}^{\ell}\left(t\right),\label{eq:c1}\\
\frac{\partial}{\partial t}\mathbf{E}_{S}^{\ell}\left(x,t\right) & = & c^{2}\boldsymbol{\nabla}\times\mathbf{B}_{S}^{\ell}\left(x,t\right)-\frac{1}{\epsilon_{0}}\mathbf{J}^{\ell}\left(x,t\right),\label{eq:ER}\\
\frac{\partial}{\partial t}\mathbf{B}_{S}^{\ell}\left(x,t\right) & = & -\boldsymbol{\nabla}\times\mathbf{E}_{S}^{\ell}\left(x,t\right).\label{eq:BR}
\end{eqnarray}
Here, the scattered field is an ensemble of classical EM field $\{\mathbf{E}_{S}^{\ell},\mathbf{B}_{S}^{\ell}\}$
with the initial condition in the vacuum (i.e. $\mathbf{E}_{S}^{\ell}\left(x,0\right)=\mathbf{B}_{S}^{\ell}\left(x,0\right)=0$).
The average current generated by the TLS is $\mathbf{J^{\ell}}\left(x,t\right)=-2\omega_{0}\text{Im}[c_{0}^{\ell}\left(t\right)c_{1}^{\ell}\left(t\right)^{*}]\mathbf{d}\left(x\right)$.
In Eqs.~(\ref{eq:c0})\textendash (\ref{eq:BR}), note that we work
with the wavefunction of the electronic system, $\left|\psi^{\ell}\right\rangle =c_{0}^{\ell}\left|0\right\rangle +c_{1}^{\ell}\left|1\right\rangle $,
and we can build the electronic density matrix as $\rho_{ij}=\frac{1}{N}\sum_{\ell}c_{i}^{\ell}c_{j}^{\ell*}$
for $N$ trajectories. The electric dipole coupling is $H_{01}\left(t\right)=\int_{-\infty}^{\infty}\mathrm{d}x[\mathbf{E}_{L}\left(x,t\right)+\mathbf{E}_{S}^{\ell}\left(x,t\right)]\cdot\mathbf{d}\left(x\right)$,
which is composed of the laser driving term (coupled to the laser
field $\mathbf{E}_{L}$) and the radiation self-interaction term (coupled
to the scattered field $\mathbf{E}_{S}^{\ell}$).

While Eqs.~(\ref{eq:c0})\textendash (\ref{eq:BR}) constitute Ehrenfest
dynamics, the augmentations needed for the ``+R'' correction are
threefold: 1.~population relaxation, 2.~stochastic dephasing, and
3.~EM field rescaling.

\emph{1.~Population relaxation.} Note that Ehrenfest dynamics do
not fully include spontaneous emission.\cite{li_mixed_2018} To recover
spontaneous emission, we add population relaxation after every time
step $dt$ propagating Eqs.~(\ref{eq:c0})\textendash (\ref{eq:BR}),
and adjust the electronic wave function by
\begin{eqnarray}
c_{0}^{\ell} & \rightarrow & \frac{c_{0}^{\ell}}{\left|c_{0}^{\ell}\right|}\sqrt{\left|c_{0}^{\ell}\right|^{2}+k_{R}\left|c_{1}^{\ell}\right|^{2}dt},\label{eq:c0'}\\
c_{1}^{\ell} & \rightarrow & \frac{c_{1}^{\ell}}{\left|c_{1}^{\ell}\right|}\sqrt{\left|c_{1}^{\ell}\right|^{2}-k_{R}\left|c_{1}^{\ell}\right|^{2}dt}.\label{eq:c1'}
\end{eqnarray}
Here, $k_{R}\equiv2\kappa\left|c_{1}^{\ell}\right|^{2}\text{Im}\left[e^{i\phi_{\ell}}c_{0}^{\ell}c_{1}^{\ell*}/\left|c_{0}^{\ell}c_{1}^{\ell*}\right|\right]^{2}$
is the +R relaxation rate ($k_{R}=0$ if $\rho_{11}=0$) and $\kappa=\mu^{2}\omega_{0}/\hbar\epsilon_{0}c$
is the Fermi\textquoteright s golden rule (FGR) rate for 1D space.
$\phi_{\ell}\in\left[0,2\pi\right]$ is a random phase chosen at the
beginning of each trajectory.

\emph{2.~Stochastic dephasing.} After each time step, we implement
dephasing by multiplication with another random phase $\Phi\in\left[0,2\pi\right]$:
\begin{equation}
c_{0}^{\ell}\rightarrow c_{0}^{\ell}e^{i\Phi}~\text{if RN}<\kappa\left|c_{1}^{\ell}\right|^{2}dt
\end{equation}
where $\text{RN}\in\left[0,1\right]$ is a random number.

\emph{3.~EM field rescaling.} In order to conserve energy, after
every population relaxation event in Eqs.~(\ref{eq:c0'})\textendash (\ref{eq:c1'}),
we rescale the scattered EM field by
\begin{eqnarray}
\mathbf{E}_{S}^{\ell} & \rightarrow & \mathbf{E}_{S}^{\ell}+\text{\ensuremath{\alpha^{\ell}}}\delta\mathbf{E}_{S},\label{eq:rescaling_operator_E}\\
\mathbf{B}_{S}^{\ell} & \rightarrow & \mathbf{B}_{S}^{\ell}+\text{\ensuremath{\beta^{\ell}}}\delta\mathbf{B}_{S}.\label{eq:rescaling_operator_B}
\end{eqnarray}
Here, the rescaling fields are chosen to be $\delta\mathbf{E}_{S}\left(x\right)=-\mu\sqrt{\frac{a}{\pi}}4a^{2}x^{2}e^{-ax^{2}}\mathbf{e}_{z}$
and $\delta\mathbf{B}_{S}\left(x\right)=\mu\sqrt{\frac{a}{\pi}}\frac{4}{3}a^{2}x^{3}e^{-ax^{2}}\mathbf{e}_{y}$,
which depend on the polarization distribution $\mathbf{d}$ only.
The coefficients $\text{\ensuremath{\alpha^{\ell}}},\beta^{\ell}$
can be calculated to be
\begin{eqnarray}
\text{\ensuremath{\alpha^{\ell}}} & = & dt\sqrt{\frac{c\omega_{0}k_{R}}{\Lambda\epsilon_{0}}\frac{\left|c_{1}^{\ell}\right|^{2}}{\int\mathrm{d}v\left|\delta\mathbf{E}_{S}\right|^{2}}}\text{sgn}\left(\text{Im}\left[c_{0}^{\ell}c_{1}^{\ell*}e^{i\phi^{\ell}}\right]\right)\label{eq:alpha}\\
\text{\ensuremath{\beta^{\ell}}} & = & dt\sqrt{\frac{c\omega_{0}k_{R}}{\Lambda}\frac{\mu_{0}\left|c_{1}^{\ell}\right|^{2}}{\int\mathrm{d}v\left|\delta\mathbf{B}_{S}\right|^{2}}}\text{sgn}\left(\text{Im}\left[c_{0}^{\ell}c_{1}^{\ell*}e^{i\phi^{\ell}}\right]\right).\label{eq:beta}
\end{eqnarray}
Note that, for a Gaussian polarization distribution in 1D, the self-interference
length is determined by $\Lambda=\frac{2}{3}\sqrt{\frac{2\pi}{a}}$,
and we use the random phase $\phi_{\ell}$ in Eqs.~(\ref{eq:alpha})\textendash (\ref{eq:beta}).
Ref.~\onlinecite{li_comparison_2019} demonstrates that Eqs.~(\ref{eq:c0'})\textendash (\ref{eq:beta})
conserve energy for a ensemble of trajectories.

Finally, all together, the Ehrenfest+R dynamics are outlined in Eqs.~(\ref{eq:c0})\textendash (\ref{eq:beta}).
One key motivation for Ehrenfest+R dynamics was the drive to distinguish
the coherent emission ($\langle\mathbf{E}_{S}\rangle^{2}$) from the
total emission intensity $\langle\mathbf{E}_{S}^{2}\rangle$. This
distinction is achieved by averaging over an ensemble of trajectories,
which has been very successful in the area of electron\textendash nuclear
dynamics, including surface hopping\cite{tully_molecular_1990} and
multiple spawning\cite{ben-nun_ab_2000} trajectories. Importantly,
the computational cost of running several Ehrenfest+R trajectories
is moderate and easy to parallelize, such that simulating a large
network of quantum emitters or an environment with complicated dielectrics
should be possible in the future.

\paragraph*{Mollow Triplet Spectrum\label{sec:Mollow-Triplet-Spectrum}}

The Mollow triplet arises when the FGR rate is much smaller than
the Rabi frequency which is in turn smaller than the energy spacing
of the TLS, i.e. $\kappa\ll\Omega_{r}\ll\omega_{0}$. For this reason,
we choose parameters as follows: $\kappa/\omega_{0}=1/800$ and $\Omega_{r}/\omega_{0}=1/7$.
The laser detuning between the incident laser field frequency and
the TLS emission frequency is denoted as $\delta_{L}=\omega_{L}-\omega_{0}$.
We assume that the initial state of the TLS is the ground state and
uncoupled to the EM field, and the entire system reaches steady state
at long times, which is independent from the initial condition. 

\begin{figure}
\begin{centering}
\includegraphics{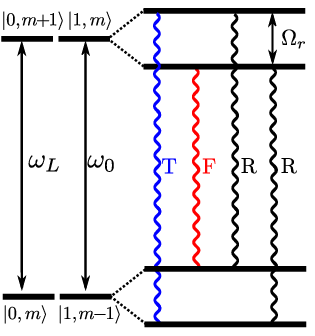}\includegraphics{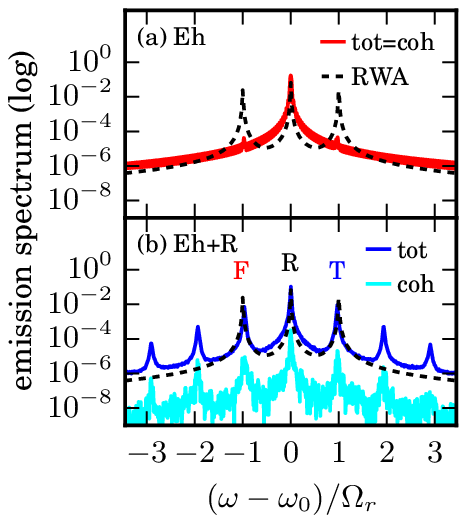}
\par\end{centering}
\caption{(Left) Energy diagram at resonance ($\omega_{L}=\omega_{0}$) according
to a quantum dressed state representation. The ground state combined
with $m+1$ incident photons $\left|0,m+1\right\rangle $ aligns energetically
with the excited state combined with $m$ photons $\left|1,m\right\rangle $
for all $m>0$. After diagonalization, two dressed states are separated
by the Rabi frequency $\left|\Omega_{r}\right|$. The radiative transitions
among the dressed states are denoted as: Rayleigh emission ($\Omega_{\mathrm{R}}=\omega_{L}$,
labeled R in black), fluorescence emission ($\Omega_{\mathrm{F}}=\omega_{L}-\Omega_{r}$,
labeled F in red), and ``three-photon'' emission ($\Omega_{\mathrm{T}}=\omega_{L}+\Omega_{r}$,
labeled T in blue). (Right) Semi-log plot of the scattering spectra
of the driven TLS as obtained by (a) Ehrenfest and (b) Ehrenfest+R
dynamics. The black dashed line is the analytical result given by
the quantum OBE within the RWA. The Ehrenfest dynamics (colored red)
yield only coherent emission and that emission is restricted to the
incident laser frequency. By contrast, for Ehrenfest+R dynamics, we
find both the coherent emission $|\langle\mathbf{E}_{S}(\omega)\rangle|^{2}$
(colored cyan) and the total emission spectrum $\langle|\mathbf{E}_{S}(\omega)|^{2}\rangle$
(colored blue). Note that the Ehrenfest+R dynamics predict higher
order emission peaks at $\omega-\omega_{0}=\pm2\Omega_{r},\pm3\Omega_{r},\cdots$.
\label{fig:Sresonant-spectrum}}
\end{figure}

We first focus on the \emph{resonant} Mollow triplet ($\delta_{L}=0$).
In Fig.~\ref{fig:Sresonant-spectrum}(a)\textendash (b), we plot
results from Ehrenfest dynamics and Ehrenfest+R dynamics. We find
that Ehrenfest+R dynamics predict a resonant Mollow triplet spectrum
at the correct three frequencies and the emission is predominantly
incoherent as expected from QED. By contrast, without the +R correction,
Ehrenfest dynamics produce only Rayleigh emission (i.e. only one frequency),
and the emission is entirely coherent. To explain this behavior, we
recall that the standard Ehrenfest approach can recover the correct
spontaneous emission rate only when the quantum subsystem is mostly
in the ground state.\cite{li_mixed_2018,li_necessary_2018} In the
presence of the strong driving field, however, the TLS quantum system
is almost always above saturation and oscillating between the ground
and excited state, so that the standard Ehrenfest dynamics cannot
emit the radiation field correctly. This result highlights the importance
of spontaneous emission and stochastic dephasing as far as recovering
key features of QED. As far as characterizing the Mollow triplet,
Ehrenfest dynamics are not enough, but Ehrenfest+R dynamics appear
sufficient.

\begin{figure*}
\begin{centering}
\textbf{\footnotesize{}(a)\hspace*{8cm}(b)\hspace*{7cm}}{\footnotesize\par}
\par\end{centering}
\begin{centering}
\includegraphics{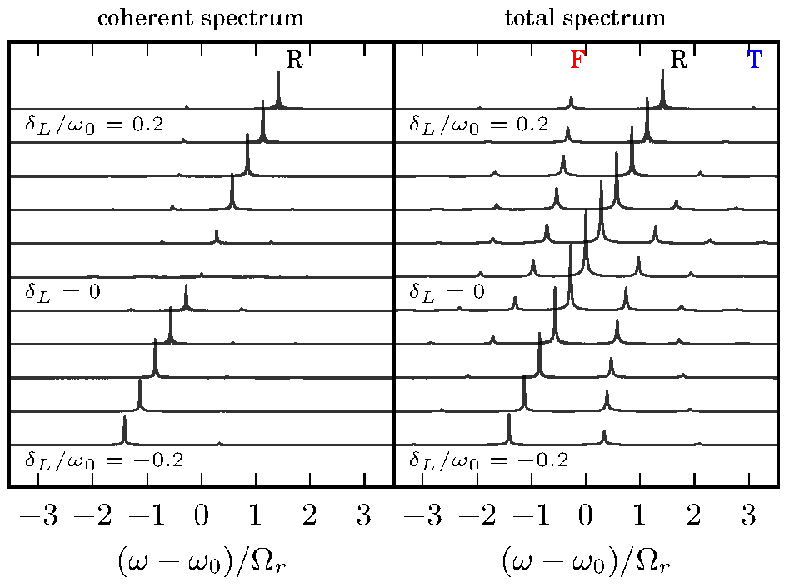}\includegraphics{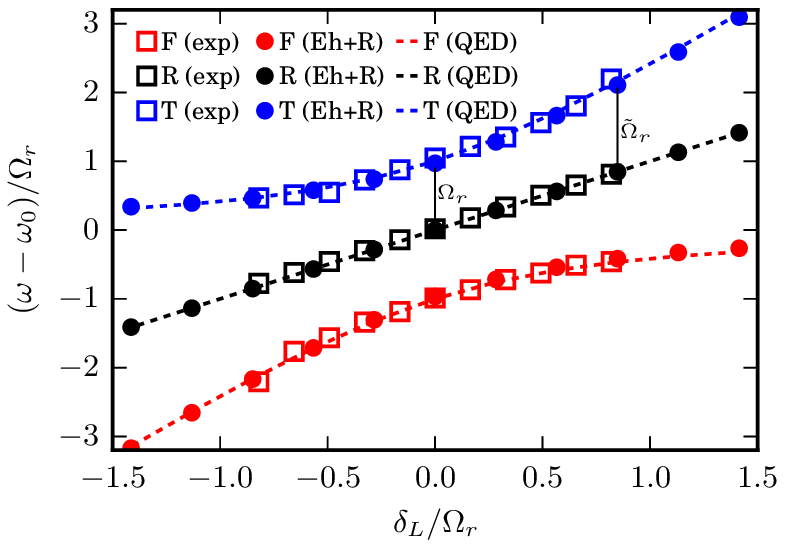}
\par\end{centering}
\caption{(a) Total scattering spectra of the driven TLS when varying the laser
detuning $\delta_{L}$. We plot both the coherent scattering spectrum
$|\langle\mathbf{E}_{S}(\omega)\rangle|^{2}$ (left) and the total
intensity spectrum $\langle|\mathbf{E}_{S}(\omega)|^{2}\rangle$ (right)
as obtained by the Ehrenfest+R dynamics. Note that coherent emission
occurs only at the Rayleigh emission ($\omega=\omega_{L}$) and vanishes
at resonance ($\delta_{L}=0$) for the case of a strong incident field
($\Omega_{r}\gg\kappa$)\ref{note1}. (b) We plot three peak positions
as a function of the laser detuning for Rayleigh (colored black),
fluorescence (colored red), and ``three-photon'' (colored blue).
The dashed lines are the quantum results, i.e. $\Omega_{\mathrm{R}}=\omega_{L}$,
$\Omega_{\mathrm{F}}=\omega_{L}-\tilde{\Omega}_{r}$, and $\Omega_{\mathrm{T}}=\omega_{L}+\tilde{\Omega}_{r}$.
The solid circles are the Ehrenfest+R data extracted from the left
panel. The boxes are experimental data from Ref.~25.
As a function of the detuning, Ehrenfest+R dynamics predict the correct
peak positions in agreement with QED and experiment.  \label{fig:detuning-dependence-spectrum}}
\end{figure*}

With this promising result in mind, we now turn our attention to
the Mollow triplet spectroscopy at varying incident field frequencies
(non-zero detuning). From the energy diagram of QED, when the incident
laser is off-resonant ($\delta_{L}\neq0$), we expect that the Rayleigh
scattering frequency changes linearly with respect to the laser detuning
($\Omega_{\mathrm{R}}=\omega_{L}=\omega_{0}+\delta_{L}$) and the
separation of the sidebands matches the generalized Rabi frequency,
$\tilde{\Omega}_{r}=\sqrt{\Omega_{r}^{2}+\delta_{L}^{2}}$, so that
$\Omega_{\mathrm{T}}=\omega_{L}+\tilde{\Omega}_{r}$ and $\Omega_{\mathrm{F}}=\omega_{L}-\tilde{\Omega}_{r}$.
In Fig.~\ref{fig:detuning-dependence-spectrum}, we plot both the
coherent spectrum ($|\langle\mathbf{E}_{S}(\omega)\rangle|^{2}$)
and the total intensity spectrum from Ehrenfest+R dynamics and extract
the peak positions as a function of the laser detuning. We show conclusively
that the Ehrenfest+R dynamics produce an accurate detuning dependence
for the Mollow triplet peak positions in agreement with QED and experimental
results.\cite{renaud_nonstationary_1977,ulhaq_cascaded_2012,ulhaq_detuning-dependent_2013,vamivakas_spin-resolved_2009}
Interestingly, as opposed to the QED data (which is based on the
quantum OBE with the RWA), the Ehrenfest+R dynamics (which go beyond
the RWA) predict higher order emission sidebands ($\omega_{L}\pm n\tilde{\Omega}_{r}$)
for $n=2,3,\cdots$ (see Fig.~\ref{fig:Sresonant-spectrum}(b)).

\begin{figure}
\begin{centering}
\par\end{centering}
\textbf{\footnotesize{}(a)\hspace*{3.5cm}(b)\hspace*{3.5cm}}{\footnotesize\par}
\begin{centering}
\includegraphics{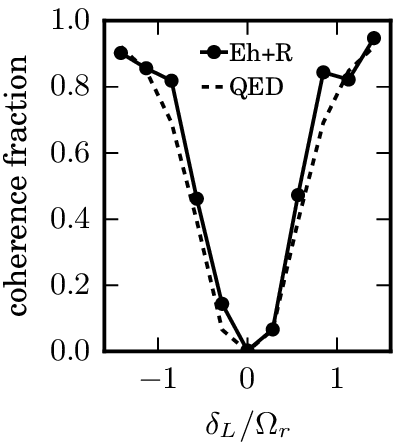}\enskip{}\includegraphics{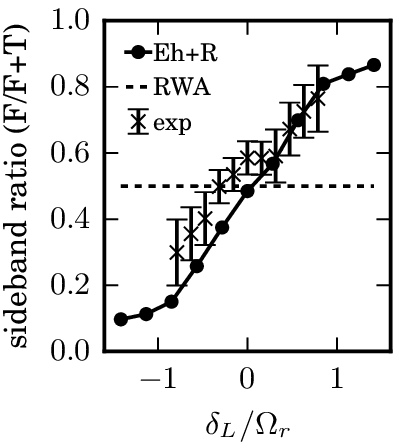}
\par\end{centering}
\caption{(a) The coherence fraction of the scattering intensity at the Rayleigh
frequency ($\omega=\Omega_{\mathrm{R}}$) as a function of the laser
detuning. The circles are the Ehrenfest+R results and the dash line
is the analytical result $|\langle\mathbf{E}_{S}(\Omega_{\mathrm{R}})\rangle|^{2}/\langle|\mathbf{E}_{S}(\Omega_{\mathrm{R}})|^{2}\rangle=4\delta_{L}^{2}\left(\Omega_{r}^{2}+\delta_{L}^{2}\right)/\left(\Omega_{r}^{2}+2\delta_{L}^{2}\right)^{2}$.
Note that the Rayleigh scattering signal is dominated by incoherent
scattering when the incident laser is on resonance ($\delta_{L}=0$)
and becomes more coherent when off-resonant. (b) The relative sideband
intensity for the fluorescence signal (denoted as $F/(F+T)$) as a
function of the detuning. Note that the Ehrenfest+R results agree
closely with the experimental data (adapted from Ref.~26).
By contrast, the RWA always predicts (incorrectly) a symmetric Mollow
triplet with $F/(F+T)=0.5$.\label{fig:ratios}}
\end{figure}

Next, let us compare the coherent emission and the total intensity.
In Fig.~\ref{fig:detuning-dependence-spectrum}, one can observe
that the EM fields emitted by Ehrenfest+R dynamics are coherent only
at the Rayleigh frequency ($\Omega_{\mathrm{R}}=\omega_{L}$) and
the sidebands are almost always incoherent.\bibnote{The fraction of the coherent component also depends on the incident
field intensity. As shown in Ref.~\onlinecite{li_comparison_2019},
the Ehrenfest+R results and analytical calculations based on the quantum
OBE are in quantitative agreement\label{note1}} Furthermore, the fraction of the coherent component at the Rayleigh
frequency strongly depends on the laser detuning and will be suppressed
when the laser frequency is at resonance. To quantify these observations,
in Fig.~\ref{fig:ratios}(a), we plot the coherence fraction $|\langle\mathbf{E}_{S}(\omega)\rangle|^{2}/\langle|\mathbf{E}_{S}(\omega)|^{2}\rangle$
for the Rayleigh peak as a function of the laser detuning and compare
against QED results.\cite{cohen-tannoudji_atom-photon_1998} To rationalize
the detuning dependence of the coherence fraction, we recall that
the spontaneous emission of a TLS quantum emitter on the excited state
is dominated by incoherent emission.\cite{chen_ehrenfest+r_2019}
Thus, when at resonance, the quantum emitter will be pumped to the
excited state by the incident laser so that, after emission, all scattered
light will be almost exclusively incoherent. By contrast, when the
laser frequency is largely off-resonant, the quantum emitter will
mostly stay near the ground state, so that coherent emission will
dominate. 

Finally, we focus on the sidebands of Mollow triplet. In Fig.~\ref{fig:ratios}(b),
we plot the sideband ratio for the fluorescence peak, $\langle|\mathbf{E}_{S}(\Omega_{\mathrm{F}})|^{2}\rangle/(\langle|\mathbf{E}_{S}(\Omega_{\mathrm{F}})|^{2}\rangle+\langle|\mathbf{E}_{S}(\Omega_{\mathrm{T}})|^{2}\rangle)$,
as a function of the laser detuning. As is clear from the figure,
Ehrenfest+R dynamics produce asymmetric sideband ratios, namely the
F peak is higher than the T peak when the incident laser has a positive
detuning ($\delta_{L}>0$) and vice versa. In fact, the Ehrenfest+R
model predicts a sideband ratio in near quantitative agreement with
experimental measurements.\cite{ulhaq_detuning-dependent_2013,vamivakas_spin-resolved_2009}
By contrast, the RWA result incorrectly predicts a symmetric Mollow
triplet spectrum for all $\delta_{L}$,which reminds us that the RWA
is valid only near resonance. The fact that the Ehrenfest+R dynamics
can capture asymmetric sidebands indicates that contributions from
the non-RWA terms (such as off-resonance and higher order\cite{browne_resonance_2000,yan_effects_2013,ge_mollow_2013,laucht_breaking_2016})
are effectively included in this semi-classical model.

\paragraph*{Discussion and Conclusions\label{sec:Discussion}}

In this letter, we have shown that, quite surprisingly, the key universal
features of the Mollow triplet can be modeled with mixed quantum\textendash classical
theory and not just QED: we require only quantum matter plus classical
light. The basic idea is that Ehrenfest+R dynamics incorporate both
electronic self-interaction (as caused by explicit propagation of
the scattered EM field) as well some of the quantum fluctuations of
light (as caused by enforcing stochastic relaxation and dephasing).
Furthermore, we have shown that because this method does not rely
on the RWA, Ehrenfest+R can recover the correct scattering spectrum
in a wide range of laser detuning far away from resonance, which is
not possible with the usual QBE within the RWA. 

Looking forward, on the one hand, even though we find accurate results
with a semiclassical model, one should be cautious about extrapolating
our results and rushing into judgment as to what is quantum and what
is classical.\cite{miller_perspective:_2012} As a precaution, we
note that the semiclassical model above cannot describe some quantum
effects of the Mollow triplet that involve higher-order correlations,
such as radiative cascaded emission and photon antibunching. On the
other hand, it will be interesting to find out exactly how far this
(or a similar) semiclassical model of light\textendash matter interactions
can take us. The success of the Ehrenfest+R dynamics in recovering
the Mollow triplet quantitatively is evidence that, with energy conservation,
several nuanced features of quantum optics can be dissected classically.
Just as a host of semiclassical non-adiabatic dynamics algorithms
have transformed the field of photochemistry and molecular electronics
over the past 20 years,\cite{fiedlschuster_surface_2017,castro_enhancing_2015,schild_exact_2017,cotton_symmetrical_2013-1}
our hope is that, with new semiclassical approaches for light\textendash matter
interactions, one will soon be able to realistically model many exciting
new collective phenomena, including resonant energy transfer,\cite{li_necessary_2018,salam_unified_2018}
superradiance,\cite{spano_superradiance_1989,lim_exciton_2004,spano_spectral_2010}
cavity effects,\cite{hoffmann_light-matter_2018,ribeiro_polariton_2018,martinez-martinez_can_2018}
and nanoplasmonics.\cite{sukharev_optics_2017} Further work in this
regard will certainly be very exciting.

\section*{Acknowledgment}

This material is based upon work supported by the U.S. Department
of Energy, Office of Science, Office of Basic Energy Sciences under
Award Number DE-SC0019397. The research of AN is supported by the
Israel-U.S. Binational Science Foundation. The authors thank Simone
Luca Portalupi and Peter Michler for providing experimental data.
\bibliographystyle{achemso}
\bibliography{MyLibrary}

\providecommand{\latin}[1]{#1}
\providecommand*\mcitethebibliography{\thebibliography}
\csname @ifundefined\endcsname{endmcitethebibliography}
  {\let\endmcitethebibliography\endthebibliography}{}
\begin{mcitethebibliography}{57}
\providecommand*\natexlab[1]{#1}
\providecommand*\mciteSetBstSublistMode[1]{}
\providecommand*\mciteSetBstMaxWidthForm[2]{}
\providecommand*\mciteBstWouldAddEndPuncttrue
  {\def\EndOfBibitem{\unskip.}}
\providecommand*\mciteBstWouldAddEndPunctfalse
  {\let\EndOfBibitem\relax}
\providecommand*\mciteSetBstMidEndSepPunct[3]{}
\providecommand*\mciteSetBstSublistLabelBeginEnd[3]{}
\providecommand*\EndOfBibitem{}
\mciteSetBstSublistMode{f}
\mciteSetBstMaxWidthForm{subitem}{(\alph{mcitesubitemcount})}
\mciteSetBstSublistLabelBeginEnd
  {\mcitemaxwidthsubitemform\space}
  {\relax}
  {\relax}

\bibitem[Newstein(1968)]{newstein_spontaneous_1968}
Newstein,~M.~C. Spontaneous {Emission} in the {Presence} of a {Prescribed}
  {Classical} {Field}. \emph{Phys. Rev.} \textbf{1968}, \emph{167},
  89--96\relax
\mciteBstWouldAddEndPuncttrue
\mciteSetBstMidEndSepPunct{\mcitedefaultmidpunct}
{\mcitedefaultendpunct}{\mcitedefaultseppunct}\relax
\EndOfBibitem
\bibitem[Mollow(1969)]{mollow_power_1969}
Mollow,~B.~R. Power {Spectrum} of {Light} {Scattered} by {Two}-{Level}
  {Systems}. \emph{Phys. Rev.} \textbf{1969}, \emph{188}, 1969--1975\relax
\mciteBstWouldAddEndPuncttrue
\mciteSetBstMidEndSepPunct{\mcitedefaultmidpunct}
{\mcitedefaultendpunct}{\mcitedefaultseppunct}\relax
\EndOfBibitem
\bibitem[Kimble and Mandel(1976)Kimble, and Mandel]{kimble_theory_1976}
Kimble,~H.~J.; Mandel,~L. Theory of {Resonance} {Fluorescence}. \emph{Phys.
  Rev. A} \textbf{1976}, \emph{13}, 2123--2144\relax
\mciteBstWouldAddEndPuncttrue
\mciteSetBstMidEndSepPunct{\mcitedefaultmidpunct}
{\mcitedefaultendpunct}{\mcitedefaultseppunct}\relax
\EndOfBibitem
\bibitem[del Valle and Laussy(2010)del Valle, and
  Laussy]{del_valle_mollow_2010}
del Valle,~E.; Laussy,~F.~P. Mollow {Triplet} under {Incoherent} {Pumping}.
  \emph{Phys. Rev. Lett.} \textbf{2010}, \emph{105}, 233601\relax
\mciteBstWouldAddEndPuncttrue
\mciteSetBstMidEndSepPunct{\mcitedefaultmidpunct}
{\mcitedefaultendpunct}{\mcitedefaultseppunct}\relax
\EndOfBibitem
\bibitem[Moelbjerg \latin{et~al.}(2012)Moelbjerg, Kaer, Lorke, and
  M{\o}rk]{moelbjerg_resonance_2012}
Moelbjerg,~A.; Kaer,~P.; Lorke,~M.; M{\o}rk,~J. Resonance {Fluorescence} from
  {Semiconductor} {Quantum} {Dots}: {Beyond} the {Mollow} {Triplet}.
  \emph{Phys. Rev. Lett.} \textbf{2012}, \emph{108}, 017401\relax
\mciteBstWouldAddEndPuncttrue
\mciteSetBstMidEndSepPunct{\mcitedefaultmidpunct}
{\mcitedefaultendpunct}{\mcitedefaultseppunct}\relax
\EndOfBibitem
\bibitem[McCutcheon and Nazir(2013)McCutcheon, and
  Nazir]{mccutcheon_model_2013}
McCutcheon,~D. P.~S.; Nazir,~A. Model of the {Optical} {Emission} of a {Driven}
  {Semiconductor} {Quantum} {Dot}: {Phonon}-{Enhanced} {Coherent} {Scattering}
  and {Off}-{Resonant} {Sideband} {Narrowing}. \emph{Phys. Rev. Lett.}
  \textbf{2013}, \emph{110}, 217401\relax
\mciteBstWouldAddEndPuncttrue
\mciteSetBstMidEndSepPunct{\mcitedefaultmidpunct}
{\mcitedefaultendpunct}{\mcitedefaultseppunct}\relax
\EndOfBibitem
\bibitem[Cohen-Tannoudji and Reynaud(1977)Cohen-Tannoudji, and
  Reynaud]{cohen-tannoudji_dressed-atom_1977}
Cohen-Tannoudji,~C.; Reynaud,~S. Dressed-atom {Description} of {Resonance}
  {Fluorescence} and {Absorption} {Spectra} of a {Multi}-level {Atom} in an
  {Intense} {Laser} {Beam}. \emph{J. Phys. B: At. Mol. Phys.} \textbf{1977},
  \emph{10}, 345--363\relax
\mciteBstWouldAddEndPuncttrue
\mciteSetBstMidEndSepPunct{\mcitedefaultmidpunct}
{\mcitedefaultendpunct}{\mcitedefaultseppunct}\relax
\EndOfBibitem
\bibitem[Ates \latin{et~al.}(2009)Ates, Ulrich, Reitzenstein, L{\"o}ffler,
  Forchel, and Michler]{ates_post-selected_2009}
Ates,~S.; Ulrich,~S.~M.; Reitzenstein,~S.; L{\"o}ffler,~A.; Forchel,~A.;
  Michler,~P. Post-{Selected} {Indistinguishable} {Photons} from the
  {Resonance} {Fluorescence} of a {Single} {Quantum} {Dot} in a {Microcavity}.
  \emph{Phys. Rev. Lett.} \textbf{2009}, \emph{103}, 167402\relax
\mciteBstWouldAddEndPuncttrue
\mciteSetBstMidEndSepPunct{\mcitedefaultmidpunct}
{\mcitedefaultendpunct}{\mcitedefaultseppunct}\relax
\EndOfBibitem
\bibitem[Hughes and Agarwal(2017)Hughes, and
  Agarwal]{hughes_anisotropy-induced_2017}
Hughes,~S.; Agarwal,~G.~S. Anisotropy-{Induced} {Quantum} {Interference} and
  {Population} {Trapping} between {Orthogonal} {Quantum} {Dot} {Exciton}
  {States} in {Semiconductor} {Cavity} {Systems}. \emph{Phys. Rev. Lett.}
  \textbf{2017}, \emph{118}, 063601\relax
\mciteBstWouldAddEndPuncttrue
\mciteSetBstMidEndSepPunct{\mcitedefaultmidpunct}
{\mcitedefaultendpunct}{\mcitedefaultseppunct}\relax
\EndOfBibitem
\bibitem[Aspect \latin{et~al.}(1980)Aspect, Roger, Reynaud, Dalibard, and
  Cohen-Tannoudji]{aspect_time_1980}
Aspect,~A.; Roger,~G.; Reynaud,~S.; Dalibard,~J.; Cohen-Tannoudji,~C. Time
  {Correlations} between the {Two} {Sidebands} of the {Resonance}
  {Fluorescence} {Triplet}. \emph{Phys. Rev. Lett.} \textbf{1980}, \emph{45},
  617--620\relax
\mciteBstWouldAddEndPuncttrue
\mciteSetBstMidEndSepPunct{\mcitedefaultmidpunct}
{\mcitedefaultendpunct}{\mcitedefaultseppunct}\relax
\EndOfBibitem
\bibitem[Schrama \latin{et~al.}(1992)Schrama, Nienhuis, Dijkerman, Steijsiger,
  and Heideman]{schrama_intensity_1992}
Schrama,~C.~A.; Nienhuis,~G.; Dijkerman,~H.~A.; Steijsiger,~C.; Heideman,~H.
  G.~M. Intensity {Correlations} {Between} the {Components} of the {Resonance}
  {Fluorescence} {Triplet}. \emph{Phys. Rev. A} \textbf{1992}, \emph{45},
  8045--8055\relax
\mciteBstWouldAddEndPuncttrue
\mciteSetBstMidEndSepPunct{\mcitedefaultmidpunct}
{\mcitedefaultendpunct}{\mcitedefaultseppunct}\relax
\EndOfBibitem
\bibitem[Nienhuis(1993)]{nienhuis_spectral_1993}
Nienhuis,~G. Spectral {Correlations} in {Resonance} {Fluorescence}. \emph{Phys.
  Rev. A} \textbf{1993}, \emph{47}, 510--518\relax
\mciteBstWouldAddEndPuncttrue
\mciteSetBstMidEndSepPunct{\mcitedefaultmidpunct}
{\mcitedefaultendpunct}{\mcitedefaultseppunct}\relax
\EndOfBibitem
\bibitem[L{\'o}pez~Carre{\~n}o \latin{et~al.}(2015)L{\'o}pez~Carre{\~n}o,
  S{\'a}nchez~Mu{\~n}oz, Sanvitto, del Valle, and
  Laussy]{lopez_carreno_exciting_2015}
L{\'o}pez~Carre{\~n}o,~J.~C.; S{\'a}nchez~Mu{\~n}oz,~C.; Sanvitto,~D.; del
  Valle,~E.; Laussy,~F.~P. Exciting {Polaritons} with {Quantum} {Light}.
  \emph{Phys. Rev. Lett.} \textbf{2015}, \emph{115}, 196402\relax
\mciteBstWouldAddEndPuncttrue
\mciteSetBstMidEndSepPunct{\mcitedefaultmidpunct}
{\mcitedefaultendpunct}{\mcitedefaultseppunct}\relax
\EndOfBibitem
\bibitem[Kimble \latin{et~al.}(1977)Kimble, Dagenais, and
  Mandel]{kimble_photon_1977}
Kimble,~H.~J.; Dagenais,~M.; Mandel,~L. Photon {Antibunching} in {Resonance}
  {Fluorescence}. \emph{Phys. Rev. Lett.} \textbf{1977}, \emph{39},
  691--695\relax
\mciteBstWouldAddEndPuncttrue
\mciteSetBstMidEndSepPunct{\mcitedefaultmidpunct}
{\mcitedefaultendpunct}{\mcitedefaultseppunct}\relax
\EndOfBibitem
\bibitem[Schuda \latin{et~al.}(1974)Schuda, Jr, and
  Hercher]{schuda_observation_1974}
Schuda,~F.; Jr,~C. R.~S.; Hercher,~M. Observation of the {Resonant} {Stark}
  {Effect} at {Optical} {Frequencies}. \emph{J. Phys. B: At. Mol. Phys.}
  \textbf{1974}, \emph{7}, L198--L202\relax
\mciteBstWouldAddEndPuncttrue
\mciteSetBstMidEndSepPunct{\mcitedefaultmidpunct}
{\mcitedefaultendpunct}{\mcitedefaultseppunct}\relax
\EndOfBibitem
\bibitem[Wu \latin{et~al.}(1975)Wu, Grove, and Ezekiel]{wu_investigation_1975}
Wu,~F.~Y.; Grove,~R.~E.; Ezekiel,~S. Investigation of the {Spectrum} of
  {Resonance} {Fluorescence} {Induced} by a {Monochromatic} {Field}.
  \emph{Phys. Rev. Lett.} \textbf{1975}, \emph{35}, 1426--1429\relax
\mciteBstWouldAddEndPuncttrue
\mciteSetBstMidEndSepPunct{\mcitedefaultmidpunct}
{\mcitedefaultendpunct}{\mcitedefaultseppunct}\relax
\EndOfBibitem
\bibitem[Wrigge \latin{et~al.}(2008)Wrigge, Gerhardt, Hwang, Zumofen, and
  Sandoghdar]{wrigge_efficient_2008}
Wrigge,~G.; Gerhardt,~I.; Hwang,~J.; Zumofen,~G.; Sandoghdar,~V. Efficient
  {Coupling} of {Photons} to a {Single} {Molecule} and the {Observation} of
  {Its} {Resonance} {Fluorescence}. \emph{Nat. Phys.} \textbf{2008}, \emph{4},
  60--66\relax
\mciteBstWouldAddEndPuncttrue
\mciteSetBstMidEndSepPunct{\mcitedefaultmidpunct}
{\mcitedefaultendpunct}{\mcitedefaultseppunct}\relax
\EndOfBibitem
\bibitem[Astafiev \latin{et~al.}(2010)Astafiev, Zagoskin, Abdumalikov, Pashkin,
  Yamamoto, Inomata, Nakamura, and Tsai]{astafiev_resonance_2010}
Astafiev,~O.; Zagoskin,~A.~M.; Abdumalikov,~A.~A.; Pashkin,~Y.~A.;
  Yamamoto,~T.; Inomata,~K.; Nakamura,~Y.; Tsai,~J.~S. Resonance {Fluorescence}
  of a {Single} {Artificial} {Atom}. \emph{Science} \textbf{2010}, \emph{327},
  840--843\relax
\mciteBstWouldAddEndPuncttrue
\mciteSetBstMidEndSepPunct{\mcitedefaultmidpunct}
{\mcitedefaultendpunct}{\mcitedefaultseppunct}\relax
\EndOfBibitem
\bibitem[Schulte \latin{et~al.}(2015)Schulte, Hansom, Jones, Matthiesen,
  Le~Gall, and Atat{\"u}re]{schulte_quadrature_2015}
Schulte,~C. H.~H.; Hansom,~J.; Jones,~A.~E.; Matthiesen,~C.; Le~Gall,~C.;
  Atat{\"u}re,~M. Quadrature {Squeezed} {Photons} from a {Two}-level {System}.
  \emph{Nature} \textbf{2015}, \emph{525}, 222--225\relax
\mciteBstWouldAddEndPuncttrue
\mciteSetBstMidEndSepPunct{\mcitedefaultmidpunct}
{\mcitedefaultendpunct}{\mcitedefaultseppunct}\relax
\EndOfBibitem
\bibitem[Xu \latin{et~al.}(2007)Xu, Sun, Berman, Steel, Bracker, Gammon, and
  Sham]{xu_coherent_2007}
Xu,~X.; Sun,~B.; Berman,~P.~R.; Steel,~D.~G.; Bracker,~A.~S.; Gammon,~D.;
  Sham,~L.~J. Coherent {Optical} {Spectroscopy} of a {Strongly} {Driven}
  {Quantum} {Dot}. \emph{Science} \textbf{2007}, \emph{317}, 929--932\relax
\mciteBstWouldAddEndPuncttrue
\mciteSetBstMidEndSepPunct{\mcitedefaultmidpunct}
{\mcitedefaultendpunct}{\mcitedefaultseppunct}\relax
\EndOfBibitem
\bibitem[Muller \latin{et~al.}(2007)Muller, Flagg, Bianucci, Wang, Deppe, Ma,
  Zhang, Salamo, Xiao, and Shih]{muller_resonance_2007}
Muller,~A.; Flagg,~E.~B.; Bianucci,~P.; Wang,~X.~Y.; Deppe,~D.~G.; Ma,~W.;
  Zhang,~J.; Salamo,~G.~J.; Xiao,~M.; Shih,~C.~K. Resonance {Fluorescence} from
  a {Coherently} {Driven} {Semiconductor} {Quantum} {Dot} in a {Cavity}.
  \emph{Phys. Rev. Lett.} \textbf{2007}, \emph{99}, 187402\relax
\mciteBstWouldAddEndPuncttrue
\mciteSetBstMidEndSepPunct{\mcitedefaultmidpunct}
{\mcitedefaultendpunct}{\mcitedefaultseppunct}\relax
\EndOfBibitem
\bibitem[Vamivakas \latin{et~al.}(2009)Vamivakas, Zhao, Lu, and
  Atat{\"u}re]{vamivakas_spin-resolved_2009}
Vamivakas,~A.~N.; Zhao,~Y.; Lu,~C.-Y.; Atat{\"u}re,~M. Spin-resolved
  {Quantum}-dot {Resonance} {Fluorescence}. \emph{Nat. Phys.} \textbf{2009},
  \emph{5}, 198--202\relax
\mciteBstWouldAddEndPuncttrue
\mciteSetBstMidEndSepPunct{\mcitedefaultmidpunct}
{\mcitedefaultendpunct}{\mcitedefaultseppunct}\relax
\EndOfBibitem
\bibitem[Flagg \latin{et~al.}(2009)Flagg, Muller, Robertson, Founta, Deppe,
  Xiao, Ma, Salamo, and Shih]{flagg_resonantly_2009}
Flagg,~E.~B.; Muller,~A.; Robertson,~J.~W.; Founta,~S.; Deppe,~D.~G.; Xiao,~M.;
  Ma,~W.; Salamo,~G.~J.; Shih,~C.~K. Resonantly {Driven} {Coherent}
  {Oscillations} in a {Solid}-state {Quantum} {Emitter}. \emph{Nat. Phys.}
  \textbf{2009}, \emph{5}, 203--207\relax
\mciteBstWouldAddEndPuncttrue
\mciteSetBstMidEndSepPunct{\mcitedefaultmidpunct}
{\mcitedefaultendpunct}{\mcitedefaultseppunct}\relax
\EndOfBibitem
\bibitem[Ulrich \latin{et~al.}(2011)Ulrich, Ates, Reitzenstein, L{\"o}ffler,
  Forchel, and Michler]{ulrich_dephasing_2011}
Ulrich,~S.~M.; Ates,~S.; Reitzenstein,~S.; L{\"o}ffler,~A.; Forchel,~A.;
  Michler,~P. Dephasing of {Triplet}-{Sideband} {Optical} {Emission} of a
  {Resonantly} {Driven} {InAs} / {GaAs} {Quantum} {Dot} inside a {Microcavity}.
  \emph{Phys. Rev. Lett.} \textbf{2011}, \emph{106}, 247402\relax
\mciteBstWouldAddEndPuncttrue
\mciteSetBstMidEndSepPunct{\mcitedefaultmidpunct}
{\mcitedefaultendpunct}{\mcitedefaultseppunct}\relax
\EndOfBibitem
\bibitem[Ulhaq \latin{et~al.}(2012)Ulhaq, Weiler, Ulrich, Ro{\ss}bach, Jetter,
  and Michler]{ulhaq_cascaded_2012}
Ulhaq,~A.; Weiler,~S.; Ulrich,~S.~M.; Ro{\ss}bach,~R.; Jetter,~M.; Michler,~P.
  Cascaded {Single}-photon {Emission} from the {Mollow} {Triplet} {Sidebands}
  of a {Quantum} {Dot}. \emph{Nat. Photonics} \textbf{2012}, \emph{6},
  238--242\relax
\mciteBstWouldAddEndPuncttrue
\mciteSetBstMidEndSepPunct{\mcitedefaultmidpunct}
{\mcitedefaultendpunct}{\mcitedefaultseppunct}\relax
\EndOfBibitem
\bibitem[Ulhaq \latin{et~al.}(2013)Ulhaq, Weiler, Roy, Ulrich, Jetter, Hughes,
  and Michler]{ulhaq_detuning-dependent_2013}
Ulhaq,~A.; Weiler,~S.; Roy,~C.; Ulrich,~S.~M.; Jetter,~M.; Hughes,~S.;
  Michler,~P. Detuning-dependent {Mollow} {Triplet} of a {Coherently}-driven
  {Single} {Quantum} {Dot}. \emph{Opt. Express} \textbf{2013}, \emph{21},
  4382--4395\relax
\mciteBstWouldAddEndPuncttrue
\mciteSetBstMidEndSepPunct{\mcitedefaultmidpunct}
{\mcitedefaultendpunct}{\mcitedefaultseppunct}\relax
\EndOfBibitem
\bibitem[Lagoudakis \latin{et~al.}(2017)Lagoudakis, Fischer, Sarmiento,
  McMahon, Radulaski, Zhang, Kelaita, Dory, M{\"u}ller, and Vu{\v
  c}kovi{\'c}]{lagoudakis_observation_2017}
Lagoudakis,~K.~G.; Fischer,~K.~A.; Sarmiento,~T.; McMahon,~P.~L.;
  Radulaski,~M.; Zhang,~J.~L.; Kelaita,~Y.; Dory,~C.; M{\"u}ller,~K.; Vu{\v
  c}kovi{\'c},~J. Observation of {Mollow} {Triplets} with {Tunable}
  {Interactions} in {Double} {Lambda} {Systems} of {Individual} {Hole} {Spins}.
  \emph{Phys. Rev. Lett.} \textbf{2017}, \emph{118}, 013602\relax
\mciteBstWouldAddEndPuncttrue
\mciteSetBstMidEndSepPunct{\mcitedefaultmidpunct}
{\mcitedefaultendpunct}{\mcitedefaultseppunct}\relax
\EndOfBibitem
\bibitem[Rohr \latin{et~al.}(2014)Rohr, Dupont-Ferrier, Pigeau, Verlot,
  Jacques, and Arcizet]{rohr_synchronizing_2014}
Rohr,~S.; Dupont-Ferrier,~E.; Pigeau,~B.; Verlot,~P.; Jacques,~V.; Arcizet,~O.
  Synchronizing the {Dynamics} of a {Single} {Nitrogen} {Vacancy} {Spin}
  {Qubit} on a {Parametrically} {Coupled} {Radio}-{Frequency} {Field} through
  {Microwave} {Dressing}. \emph{Phys. Rev. Lett.} \textbf{2014}, \emph{112},
  010502\relax
\mciteBstWouldAddEndPuncttrue
\mciteSetBstMidEndSepPunct{\mcitedefaultmidpunct}
{\mcitedefaultendpunct}{\mcitedefaultseppunct}\relax
\EndOfBibitem
\bibitem[Cohen-Tannoudji \latin{et~al.}(1997)Cohen-Tannoudji, Dupont-Roc, and
  Grynberg]{cohen-tannoudji_photons_1997}
Cohen-Tannoudji,~C.; Dupont-Roc,~J.; Grynberg,~G. \emph{Photons and {Atoms}:
  {Introduction} to {Quantum} {Electrodynamics}}; Wiley: New York, 1997\relax
\mciteBstWouldAddEndPuncttrue
\mciteSetBstMidEndSepPunct{\mcitedefaultmidpunct}
{\mcitedefaultendpunct}{\mcitedefaultseppunct}\relax
\EndOfBibitem
\bibitem[Cohen-Tannoudji \latin{et~al.}(1998)Cohen-Tannoudji, Dupont-Roc, and
  Grynberg]{cohen-tannoudji_atom-photon_1998}
Cohen-Tannoudji,~C.; Dupont-Roc,~J.; Grynberg,~G. \emph{Atom-{Photon}
  {Interactions}: {Basic} {Processes} and {Applications}}; Wiley: New York,
  1998\relax
\mciteBstWouldAddEndPuncttrue
\mciteSetBstMidEndSepPunct{\mcitedefaultmidpunct}
{\mcitedefaultendpunct}{\mcitedefaultseppunct}\relax
\EndOfBibitem
\bibitem[Salam(2010)]{salam_molecular_2010}
Salam,~A. \emph{Molecular {Quantum} {Electrodynamics}: {Long}-{Range}
  {Intermolecular} {Interactions}}; John Wiley \& Sons, Ltd: Hoboken, New
  Jersey, 2010\relax
\mciteBstWouldAddEndPuncttrue
\mciteSetBstMidEndSepPunct{\mcitedefaultmidpunct}
{\mcitedefaultendpunct}{\mcitedefaultseppunct}\relax
\EndOfBibitem
\bibitem[Chen \latin{et~al.}(2019)Chen, Li, Sukharev, Nitzan, and
  Subotnik]{chen_ehrenfest+r_2019}
Chen,~H.-T.; Li,~T.~E.; Sukharev,~M.; Nitzan,~A.; Subotnik,~J.~E. Ehrenfest+{R}
  dynamics. {I}. {A} mixed quantum{\textendash}classical electrodynamics
  simulation of spontaneous emission. \emph{J. Chem. Phys.} \textbf{2019},
  \emph{150}, 044102\relax
\mciteBstWouldAddEndPuncttrue
\mciteSetBstMidEndSepPunct{\mcitedefaultmidpunct}
{\mcitedefaultendpunct}{\mcitedefaultseppunct}\relax
\EndOfBibitem
\bibitem[Li \latin{et~al.}(2018)Li, Nitzan, Sukharev, Martinez, Chen, and
  Subotnik]{li_mixed_2018}
Li,~T.~E.; Nitzan,~A.; Sukharev,~M.; Martinez,~T.; Chen,~H.-T.; Subotnik,~J.~E.
  Mixed quantum-classical electrodynamics: {Understanding} spontaneous decay
  and zero-point energy. \emph{Phys. Rev. A} \textbf{2018}, \emph{97},
  032105\relax
\mciteBstWouldAddEndPuncttrue
\mciteSetBstMidEndSepPunct{\mcitedefaultmidpunct}
{\mcitedefaultendpunct}{\mcitedefaultseppunct}\relax
\EndOfBibitem
\bibitem[Li \latin{et~al.}(2019)Li, Chen, and Subotnik]{li_comparison_2019}
Li,~T.~E.; Chen,~H.-T.; Subotnik,~J.~E. Comparison of {Different} {Classical},
  {Semiclassical}, and {Quantum} {Treatments} of {Light}-{Matter}
  {Interactions}: {Understanding} {Energy} {Conservation}. \emph{J. Chem.
  Theory Comput.} \textbf{2019}, Article ASAP, arXiv: 1812.03265\relax
\mciteBstWouldAddEndPuncttrue
\mciteSetBstMidEndSepPunct{\mcitedefaultmidpunct}
{\mcitedefaultendpunct}{\mcitedefaultseppunct}\relax
\EndOfBibitem
\bibitem[Tully(1990)]{tully_molecular_1990}
Tully,~J.~C. Molecular {Dynamics} with {Electronic} {Transitions}. \emph{J.
  Chem. Phys.} \textbf{1990}, \emph{93}, 1061--1071\relax
\mciteBstWouldAddEndPuncttrue
\mciteSetBstMidEndSepPunct{\mcitedefaultmidpunct}
{\mcitedefaultendpunct}{\mcitedefaultseppunct}\relax
\EndOfBibitem
\bibitem[Ben-Nun \latin{et~al.}(2000)Ben-Nun, Quenneville, and
  Mart{\'i}nez]{ben-nun_ab_2000}
Ben-Nun,~M.; Quenneville,~J.; Mart{\'i}nez,~T.~J. Ab {Initio} {Multiple}
  {Spawning}: {Photochemistry} from {First} {Principles} {Quantum} {Molecular}
  {Dynamics}. \emph{J. Phys. Chem. A} \textbf{2000}, \emph{104},
  5161--5175\relax
\mciteBstWouldAddEndPuncttrue
\mciteSetBstMidEndSepPunct{\mcitedefaultmidpunct}
{\mcitedefaultendpunct}{\mcitedefaultseppunct}\relax
\EndOfBibitem
\bibitem[Li \latin{et~al.}(2018)Li, Chen, Nitzan, Sukharev, and
  Subotnik]{li_necessary_2018}
Li,~T.~E.; Chen,~H.-T.; Nitzan,~A.; Sukharev,~M.; Subotnik,~J.~E. A {Necessary}
  {Trade}-off for {Semiclassical} {Electrodynamics}: {Accurate} {Short}-{Range}
  {Coulomb} {Interactions} versus the {Enforcement} of {Causality}? \emph{J.
  Phys. Chem. Lett.} \textbf{2018}, \emph{9}, 5955--5961\relax
\mciteBstWouldAddEndPuncttrue
\mciteSetBstMidEndSepPunct{\mcitedefaultmidpunct}
{\mcitedefaultendpunct}{\mcitedefaultseppunct}\relax
\EndOfBibitem
\bibitem[Renaud \latin{et~al.}(1977)Renaud, Whitley, and
  Stroud]{renaud_nonstationary_1977}
Renaud,~B.; Whitley,~R.~M.; Stroud,~C.~R. Nonstationary {Two}-level {Resonance}
  {Fluorescence}. \emph{J. Phys. B: At. Mol. Phys.} \textbf{1977}, \emph{10},
  19--35\relax
\mciteBstWouldAddEndPuncttrue
\mciteSetBstMidEndSepPunct{\mcitedefaultmidpunct}
{\mcitedefaultendpunct}{\mcitedefaultseppunct}\relax
\EndOfBibitem
\bibitem[Not()]{Note-1}
The fraction of the coherent component also depends on the incident field
  intensity. As shown in Ref.~\onlinecite{li_comparison_2019}, the Ehrenfest+R
  results and analytical calculations based on the quantum OBE are in
  quantitative agreement\label{note1}\relax
\mciteBstWouldAddEndPuncttrue
\mciteSetBstMidEndSepPunct{\mcitedefaultmidpunct}
{\mcitedefaultendpunct}{\mcitedefaultseppunct}\relax
\EndOfBibitem
\bibitem[Browne and Keitel(2000)Browne, and Keitel]{browne_resonance_2000}
Browne,~D.~E.; Keitel,~C.~H. Resonance {Fluorescence} in {Intense} {Laser}
  {Fields}. \emph{J. Mod. Opt.} \textbf{2000}, \emph{47}, 1307--1337\relax
\mciteBstWouldAddEndPuncttrue
\mciteSetBstMidEndSepPunct{\mcitedefaultmidpunct}
{\mcitedefaultendpunct}{\mcitedefaultseppunct}\relax
\EndOfBibitem
\bibitem[Yan \latin{et~al.}(2013)Yan, L{\"u}, and Zheng]{yan_effects_2013}
Yan,~Y.; L{\"u},~Z.; Zheng,~H. Effects of {Counter}-rotating-wave {Terms} of
  the {Driving} {Field} on the {Spectrum} of {Resonance} {Fluorescence}.
  \emph{Phys. Rev. A} \textbf{2013}, \emph{88}, 053821\relax
\mciteBstWouldAddEndPuncttrue
\mciteSetBstMidEndSepPunct{\mcitedefaultmidpunct}
{\mcitedefaultendpunct}{\mcitedefaultseppunct}\relax
\EndOfBibitem
\bibitem[Ge \latin{et~al.}(2013)Ge, Weiler, Ulhaq, Ulrich, Jetter, Michler, and
  Hughes]{ge_mollow_2013}
Ge,~R.-C.; Weiler,~S.; Ulhaq,~A.; Ulrich,~S.~M.; Jetter,~M.; Michler,~P.;
  Hughes,~S. Mollow {Quintuplets} from {Coherently} {Excited} {Quantum} {Dots}.
  \emph{Opt. Lett.} \textbf{2013}, \emph{38}, 1691--1693\relax
\mciteBstWouldAddEndPuncttrue
\mciteSetBstMidEndSepPunct{\mcitedefaultmidpunct}
{\mcitedefaultendpunct}{\mcitedefaultseppunct}\relax
\EndOfBibitem
\bibitem[Laucht \latin{et~al.}(2016)Laucht, Simmons, Kalra, Tosi, Dehollain,
  Muhonen, Freer, Hudson, Itoh, Jamieson, McCallum, Dzurak, and
  Morello]{laucht_breaking_2016}
Laucht,~A.; Simmons,~S.; Kalra,~R.; Tosi,~G.; Dehollain,~J.~P.; Muhonen,~J.~T.;
  Freer,~S.; Hudson,~F.~E.; Itoh,~K.~M.; Jamieson,~D.~N.; McCallum,~J.~C.;
  Dzurak,~A.~S.; Morello,~A. Breaking the {Rotating} {Wave} {Approximation} for
  a {Strongly} {Driven} {Dressed} {Single}-electron {Spin}. \emph{Phys. Rev. B}
  \textbf{2016}, \emph{94}, 161302\relax
\mciteBstWouldAddEndPuncttrue
\mciteSetBstMidEndSepPunct{\mcitedefaultmidpunct}
{\mcitedefaultendpunct}{\mcitedefaultseppunct}\relax
\EndOfBibitem
\bibitem[Miller(2012)]{miller_perspective:_2012}
Miller,~W.~H. Perspective: {Quantum} or {Classical} {Coherence}? \emph{J. Chem.
  Phys.} \textbf{2012}, \emph{136}, 210901\relax
\mciteBstWouldAddEndPuncttrue
\mciteSetBstMidEndSepPunct{\mcitedefaultmidpunct}
{\mcitedefaultendpunct}{\mcitedefaultseppunct}\relax
\EndOfBibitem
\bibitem[Fiedlschuster \latin{et~al.}(2017)Fiedlschuster, Handt, Gross, and
  Schmidt]{fiedlschuster_surface_2017}
Fiedlschuster,~T.; Handt,~J.; Gross,~E. K.~U.; Schmidt,~R. Surface {Hopping} in
  {Laser}-driven {Molecular} {Dynamics}. \emph{Phys. Rev. A} \textbf{2017},
  \emph{95}, 063424\relax
\mciteBstWouldAddEndPuncttrue
\mciteSetBstMidEndSepPunct{\mcitedefaultmidpunct}
{\mcitedefaultendpunct}{\mcitedefaultseppunct}\relax
\EndOfBibitem
\bibitem[Castro \latin{et~al.}(2015)Castro, Rubio, and
  Gross]{castro_enhancing_2015}
Castro,~A.; Rubio,~A.; Gross,~E. K.~U. Enhancing and {Controlling}
  {Single}-atom {High}-harmonic {Generation} {Spectra}: {A} {Time}-dependent
  {Density}-functional {Scheme}. \emph{Eur. Phys. J. B} \textbf{2015},
  \emph{88}, 191\relax
\mciteBstWouldAddEndPuncttrue
\mciteSetBstMidEndSepPunct{\mcitedefaultmidpunct}
{\mcitedefaultendpunct}{\mcitedefaultseppunct}\relax
\EndOfBibitem
\bibitem[Schild and Gross(2017)Schild, and Gross]{schild_exact_2017}
Schild,~A.; Gross,~E. K.~U. Exact {Single}-{Electron} {Approach} to the
  {Dynamics} of {Molecules} in {Strong} {Laser} {Fields}. \emph{Phys. Rev.
  Lett.} \textbf{2017}, \emph{118}, 163202\relax
\mciteBstWouldAddEndPuncttrue
\mciteSetBstMidEndSepPunct{\mcitedefaultmidpunct}
{\mcitedefaultendpunct}{\mcitedefaultseppunct}\relax
\EndOfBibitem
\bibitem[Cotton and Miller(2013)Cotton, and Miller]{cotton_symmetrical_2013-1}
Cotton,~S.~J.; Miller,~W.~H. Symmetrical {Windowing} for {Quantum} {States} in
  {Quasi}-classical {Trajectory} {Simulations}: {Application} to
  {Electronically} {Non}-adiabatic {Processes}. \emph{J. Chem. Phys.}
  \textbf{2013}, \emph{139}, 234112\relax
\mciteBstWouldAddEndPuncttrue
\mciteSetBstMidEndSepPunct{\mcitedefaultmidpunct}
{\mcitedefaultendpunct}{\mcitedefaultseppunct}\relax
\EndOfBibitem
\bibitem[Salam(2018)]{salam_unified_2018}
Salam,~A. The {Unified} {Theory} of {Resonance} {Energy} {Transfer} {According}
  to {Molecular} {Quantum} {Electrodynamics}. \emph{Atoms} \textbf{2018},
  \emph{6}, 56\relax
\mciteBstWouldAddEndPuncttrue
\mciteSetBstMidEndSepPunct{\mcitedefaultmidpunct}
{\mcitedefaultendpunct}{\mcitedefaultseppunct}\relax
\EndOfBibitem
\bibitem[Spano and Mukamel(1989)Spano, and Mukamel]{spano_superradiance_1989}
Spano,~F.~C.; Mukamel,~S. Superradiance in {Molecular} {Aggregates}. \emph{J.
  Chem. Phys.} \textbf{1989}, \emph{91}, 683--700\relax
\mciteBstWouldAddEndPuncttrue
\mciteSetBstMidEndSepPunct{\mcitedefaultmidpunct}
{\mcitedefaultendpunct}{\mcitedefaultseppunct}\relax
\EndOfBibitem
\bibitem[Lim \latin{et~al.}(2004)Lim, Bjorklund, Spano, and
  Bardeen]{lim_exciton_2004}
Lim,~S.-H.; Bjorklund,~T.~G.; Spano,~F.~C.; Bardeen,~C.~J. Exciton
  {Delocalization} and {Superradiance} in {Tetracene} {Thin} {Films} and
  {Nanoaggregates}. \emph{Phys. Rev. Lett.} \textbf{2004}, \emph{92},
  107402\relax
\mciteBstWouldAddEndPuncttrue
\mciteSetBstMidEndSepPunct{\mcitedefaultmidpunct}
{\mcitedefaultendpunct}{\mcitedefaultseppunct}\relax
\EndOfBibitem
\bibitem[Spano(2010)]{spano_spectral_2010}
Spano,~F.~C. The {Spectral} {Signatures} of {Frenkel} {Polarons} in {H}- and
  {J}-{Aggregates}. \emph{Acc. Chem. Res.} \textbf{2010}, \emph{43},
  429--439\relax
\mciteBstWouldAddEndPuncttrue
\mciteSetBstMidEndSepPunct{\mcitedefaultmidpunct}
{\mcitedefaultendpunct}{\mcitedefaultseppunct}\relax
\EndOfBibitem
\bibitem[Hoffmann \latin{et~al.}(2018)Hoffmann, Appel, Rubio, and
  Maitra]{hoffmann_light-matter_2018}
Hoffmann,~N.~M.; Appel,~H.; Rubio,~A.; Maitra,~N.~T. Light-matter
  {Interactions} via the {Exact} {Factorization} {Approach}. \emph{Eur. Phys.
  J. B} \textbf{2018}, \emph{91}, 180\relax
\mciteBstWouldAddEndPuncttrue
\mciteSetBstMidEndSepPunct{\mcitedefaultmidpunct}
{\mcitedefaultendpunct}{\mcitedefaultseppunct}\relax
\EndOfBibitem
\bibitem[Ribeiro \latin{et~al.}(2018)Ribeiro, Mart{\'i}nez-Mart{\'i}nez, Du,
  Campos-Gonzalez-Angulo, and Yuen-Zhou]{ribeiro_polariton_2018}
Ribeiro,~R.~F.; Mart{\'i}nez-Mart{\'i}nez,~L.~A.; Du,~M.;
  Campos-Gonzalez-Angulo,~J.; Yuen-Zhou,~J. Polariton {Chemistry}:
  {Controlling} {Molecular} {Dynamics} with {Optical} {Cavities}. \emph{Chem.
  Sci.} \textbf{2018}, \emph{9}, 6325--6339\relax
\mciteBstWouldAddEndPuncttrue
\mciteSetBstMidEndSepPunct{\mcitedefaultmidpunct}
{\mcitedefaultendpunct}{\mcitedefaultseppunct}\relax
\EndOfBibitem
\bibitem[Mart{\'i}nez-Mart{\'i}nez
  \latin{et~al.}(2018)Mart{\'i}nez-Mart{\'i}nez, Ribeiro,
  Campos-Gonz{\'a}lez-Angulo, and Yuen-Zhou]{martinez-martinez_can_2018}
Mart{\'i}nez-Mart{\'i}nez,~L.~A.; Ribeiro,~R.~F.;
  Campos-Gonz{\'a}lez-Angulo,~J.; Yuen-Zhou,~J. Can {Ultrastrong} {Coupling}
  {Change} {Ground}-{State} {Chemical} {Reactions}? \emph{ACS Photonics}
  \textbf{2018}, \emph{5}, 167--176\relax
\mciteBstWouldAddEndPuncttrue
\mciteSetBstMidEndSepPunct{\mcitedefaultmidpunct}
{\mcitedefaultendpunct}{\mcitedefaultseppunct}\relax
\EndOfBibitem
\bibitem[Sukharev and Nitzan(2017)Sukharev, and Nitzan]{sukharev_optics_2017}
Sukharev,~M.; Nitzan,~A. Optics of {Exciton}-plasmon {Nanomaterials}. \emph{J.
  Phys.: Condens. Matter} \textbf{2017}, \emph{29}, 443003\relax
\mciteBstWouldAddEndPuncttrue
\mciteSetBstMidEndSepPunct{\mcitedefaultmidpunct}
{\mcitedefaultendpunct}{\mcitedefaultseppunct}\relax
\EndOfBibitem
\end{mcitethebibliography}

\end{document}